\documentclass[aps,prl,twocolumn,showpacs,superscriptaddress,floatfix,10pt]{revtex4-1}
\usepackage{graphicx}
\usepackage{dcolumn}
\usepackage{bm}
\usepackage{xcolor}
\usepackage{amsfonts}
\usepackage{amssymb}
\usepackage{amsmath}
\usepackage[colorlinks,linkcolor=blue]{hyperref}

\begin{document}

\title{Dissipative dynamics in quasi-fission}

\author{V.E. Oberacker}
\author{A.S. Umar}
\affiliation{Department of Physics and Astronomy, Vanderbilt University, Nashville, Tennessee 37235, USA}
\author{C. Simenel}
\affiliation{Department of Nuclear Physics, RSPE, Australian National University, Canberra, ACT 0200, Australia}
\date{\today}

\begin{abstract}
Quasi-fission is the primary reaction mechanism that prevents the formation of
superheavy elements in heavy-ion fusion experiments. Employing the time-dependent
density functional theory approach  we study quasi-fission in the systems $^{40,48}$Ca+$^{238}$U.
Results show that for $^{48}$Ca projectiles the quasi-fission is substantially reduced in comparison to
the $^{40}$Ca case. This 
partly explains the success of superheavy element formation with $^{48}$Ca beams. 
For the first time, we also calculate 
the repartition of 
excitation energies of the two fragments in a dynamic microscopic
theory. 
The system is found in quasi-thermal equilibrium only for reactions with $^{40}$Ca.
The differences between both systems are interpreted in terms of initial neutron to proton asymmetry of the colliding partners. 
\end{abstract}
\pacs{21.60.-n,21.60.Jz}
\maketitle

The search for new elements is one of the most novel and challenging research
areas of nuclear physics. 
The discovery of a region of the nuclear chart that can sustain the so
called \textit{superheavy elements} (SHE) has lead to intense experimental activity
resulting in the discovery of elements with atomic numbers as large as
$Z=117$~\cite{oganessian2010,oganessian2012,hinde2014}.
The theoretically predicted \textit{island of stability} 
in the SHE region of the nuclear chart 
is the result of new
proton and neutron shell-closures, whose location is not precisely
known~~\cite{bender1999,staszczak2013,cwiok2005}. 
The experiments to discover these new elements are notoriously difficult, with
production cross-sections in pico-barns.
Of primary importance for the experimental investigations appear to be the choice
of target-projectile combinations that have the highest probability for forming
a compound nucleus that results in the production of the desired element.
Experimentally, two approaches have been used for the synthesis of these elements,
one utilizing doubly-magic $^{208}$Pb targets or $^{209}$Bi (cold-fusion)~\cite{hofmann2000,hofmann2002}, the other
utilizing deformed actinide targets with neutron-rich projectiles (hot-fusion), such as $^{48}$Ca~\cite{oganessian2010,oganessian2012,hinde2014}.
While both methods have been successful in synthesizing new elements the evaporation
residue cross-sections for hot-fusion were found to be several
orders of magnitude larger than those for cold fusion.
To pinpoint the root of this difference it is important to understand the details
of the reaction dynamics of these systems.
For light and medium mass systems the capture cross-section may be considered
to be the same as that for complete fusion.
For heavy systems leading to superheavy formations
however, the formation of a compound nucleus 
is dramatically reduced due to the
 quasi-fission (QF) 
process~\cite{sahm1984,schmidt1991}.
Consequently, quasi-fission is the primary reaction mechanism that limits the formation of
superheavy nuclei. Quasi-fission is characterized by nuclear contact-times that are
usually greater than 5~zs but much shorter than typical fusion-fission times which require the formation of a compound nucleus~\cite{bock1982,toke1985,shen1987,rietz2013}. 

Many experimental studies have been performed to understand the mechanisms at play in the quasi-fission process since its discovery~\cite{bock1982,toke1985,shen1987,hinde1995,itkis2004,nishio2008,nishio2012,rietz2013,kozulin2014}. 
Various theoretical models ~\cite{adamian2003,zagrebaev2007,aritomo2009} 
 have also been developed to help in the interpretation of these experimental data.
These models 
are often based on statistical or transport theories.
In this 
letter, we consider another formalism based on a many-body quantum approach.
We study quasi-fission with 
the fully microscopic time-dependent Hartree-Fock (TDHF) theory 
proposed by Dirac~\cite{dirac1930}.
The TDHF theory provides a useful foundation for a
fully microscopic many-body theory of large amplitude collective
motion. 
This approach has been widely applied to study heavy-ion collisions in nuclear physics
\cite{negele1982,simenel2012}. 
The TDHF time-evolution can correctly account for the heavy-ion interaction barriers 
\cite{washiyama2008,simenel2013b,guo2012,umar2010a} 
and thus reproduce the capture cross-sections 
in heavy systems such as
$^{48}$Ca+$^{238}$U 
\cite{umar2010a}.
It is also able to describe 
transfer and deep-inelastic reactions~\cite{umar2008a,golabek2009,kedziora2010,simenel2010,simenel2011,sekizawa2013,scamps2013,lacroix2014} 
as well as the dynamics of fission fragments~\cite{simenel2014a}.
It is therefore a tool of choice to investigate quasi-fission mechanisms.
However, the feasibility of using TDHF for quasi-fission has only been
recognized recently~\cite{wakhle2014}.
These applications have been made possible thanks to considerable improvements of computational power in the past decade.
Modern 
TDHF calculations 
are performed 
on a three-dimensional (3D) Cartesian grid with no symmetry restrictions
and with much more accurate numerical methods~\cite{bottcher1989,reinhard1988,umar1989,umar1991a,umar2005a,umar2006c,maruhn2014}.

\begin{figure}[!htb]
	\includegraphics*[width=8.6cm]{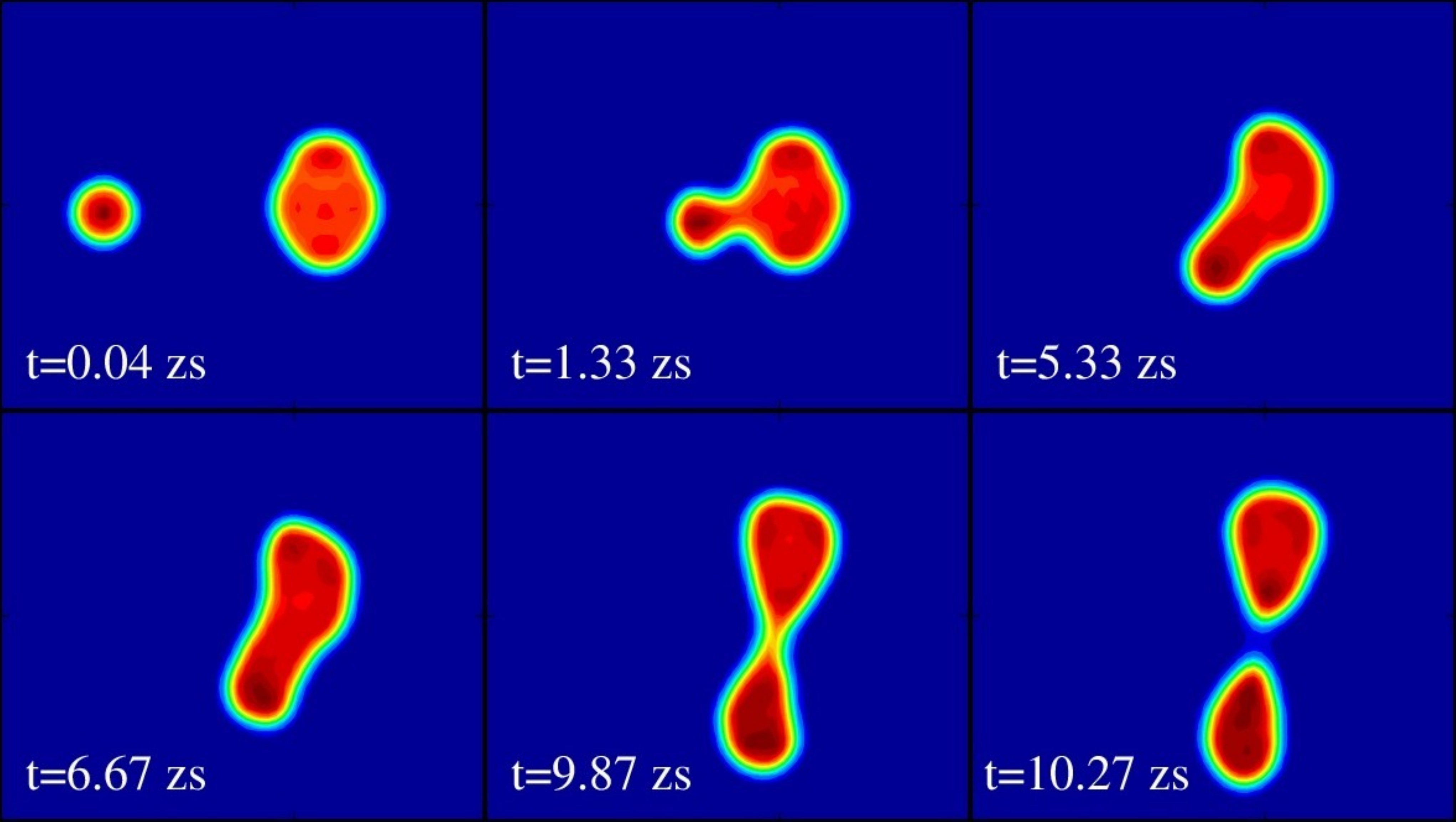}
	\caption{\protect(Color online) Quasi-fission in the reaction $^{40}$Ca+$^{238}$U
		at $E_{\mathrm{c.m.}}=209$~MeV with impact parameter $b=1.103$~fm ($L=20$).
		Shown is a contour plot of the time evolution of the mass density.
	}
	\label{fig:dens1}
\end{figure}

In the present TDHF calculations we use the Skyrme SLy4d 
energy density functional (EDF) 
\cite{kim1997}
including all of the relevant time-odd terms in the mean-field Hamiltonian.
First we generate very accurate static HF wave functions for the two nuclei on the
3D grid.
The initial separation of the two nuclei is $30$~fm. In the second
step, we apply a boost operator to the single-particle wave functions. The time-propagation
is carried out using a Taylor series expansion (up to orders $10-12$) of the
unitary mean-field propagator, with a time step $\Delta t = 0.4$~fm/c.
Let us first focus on collisions of $^{40}$Ca+$^{238}$U.
An example of TDHF calculation for this reaction at $E_{\mathrm{c.m.}}=209$~MeV
and an average orbital angular momentum quantum number $L=20$
is shown 
in Fig.~\ref{fig:dens1} 
where 
contour plots of the mass density
are plotted 
at various times. 
In this case, the 3D lattice
spans $(66 \times 56 \times 30)$~fm.
As the nuclei approach each other, a neck forms between the
two fragments which grows in size as the system begins to rotate. 
Due to the 
Coulomb repulsion and 
centrifugal forces,
the dinuclear system elongates and forms a very long neck which eventually
ruptures leading to two separated fragments.
The $^{238}$U nucleus exhibits 
a strong quadrupole 
deformation. 
In the present study, its symmetry axis was
oriented initially at $90^{\circ}$ to the internuclear axis.
For small impact parameters, this leads essentially to collisions with the side of $^{238}$U.
This orientation 
is also the one which 
leads to the largest ''contact time'' in central collisions~\cite{simenel2012,wakhle2014}.
We define the contact time as the time interval between the time $t_1$
when the two nuclear surfaces 
(defined as isodensities with half the saturation density $\rho_0/2=0.08$~fm$^{-3}$) 
first touch and
the time $t_2$ when the dinuclear system splits up again. 
In the collision shown in Fig.~\ref{fig:dens1}, 
we find
a contact time $\Delta t = 9.35$~zs 
and substantial mass transfer (66 nucleons
to the light fragment). 
This contact time and mass transfer is characteristic for QF~\cite{toke1985,rietz2013}. 
Collisions with the tip of $^{238}$U may 
also result in QF, however 
with smaller mass transfer.
In addition, the latter orientations are never found to lead to fusion in TDHF calculations~\cite{simenel2012,wakhle2014},
which is consistent with experimental observation that fusion essentially occurs in collisions with the side of the deformed actinide target~\cite{hinde1995}. 
In this letter, our goal is to investigate QF reactions in competition with the formation of a compound nucleus by fusion. 
Therefore, we investigate only  collisions with the side of $^{238}$U.

\begin{figure}[!htb]
	\includegraphics*[width=8.6cm]{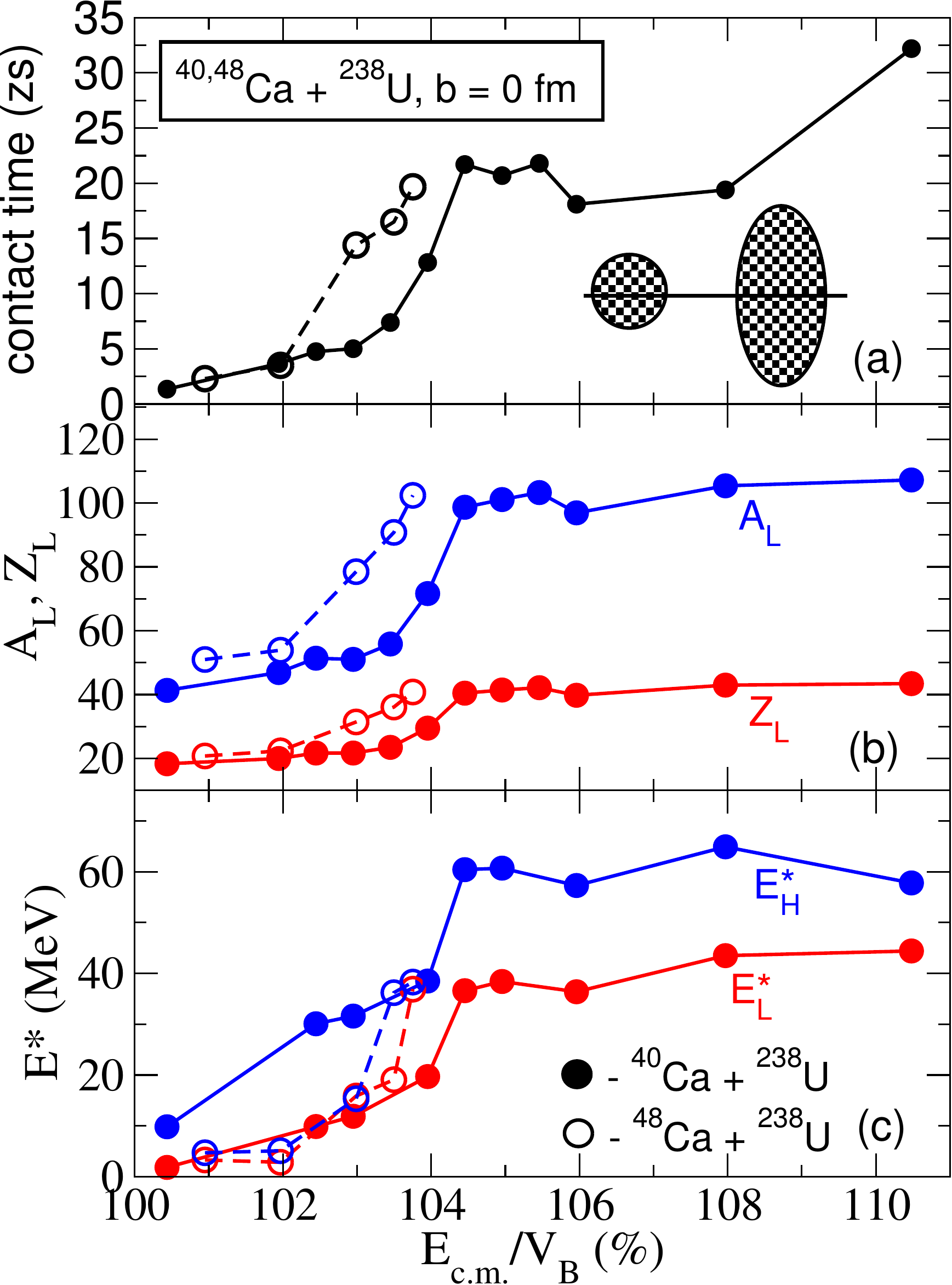}
	\caption{\protect(Color online) Several observables as a function of $E_{c.m.}/V_B$ 
		for $^{40,48}$Ca+$^{238}$U central collisions with the side of $^{238}$U. 
		The frozen HF barrier for these configurations are $V_B=199.13$~MeV with $^{40}$Ca and 
		$V_B=196.14$~MeV with $^{48}$Ca.
		(a) contact time, (b) mass and charge of the light fragment, and
		(c) excitation energy of the light and heavy fragments.
	}
	\label{fig:Ca_stack_1}
\end{figure}

\begin{figure}[!htb]
	\includegraphics*[width=8.6cm]{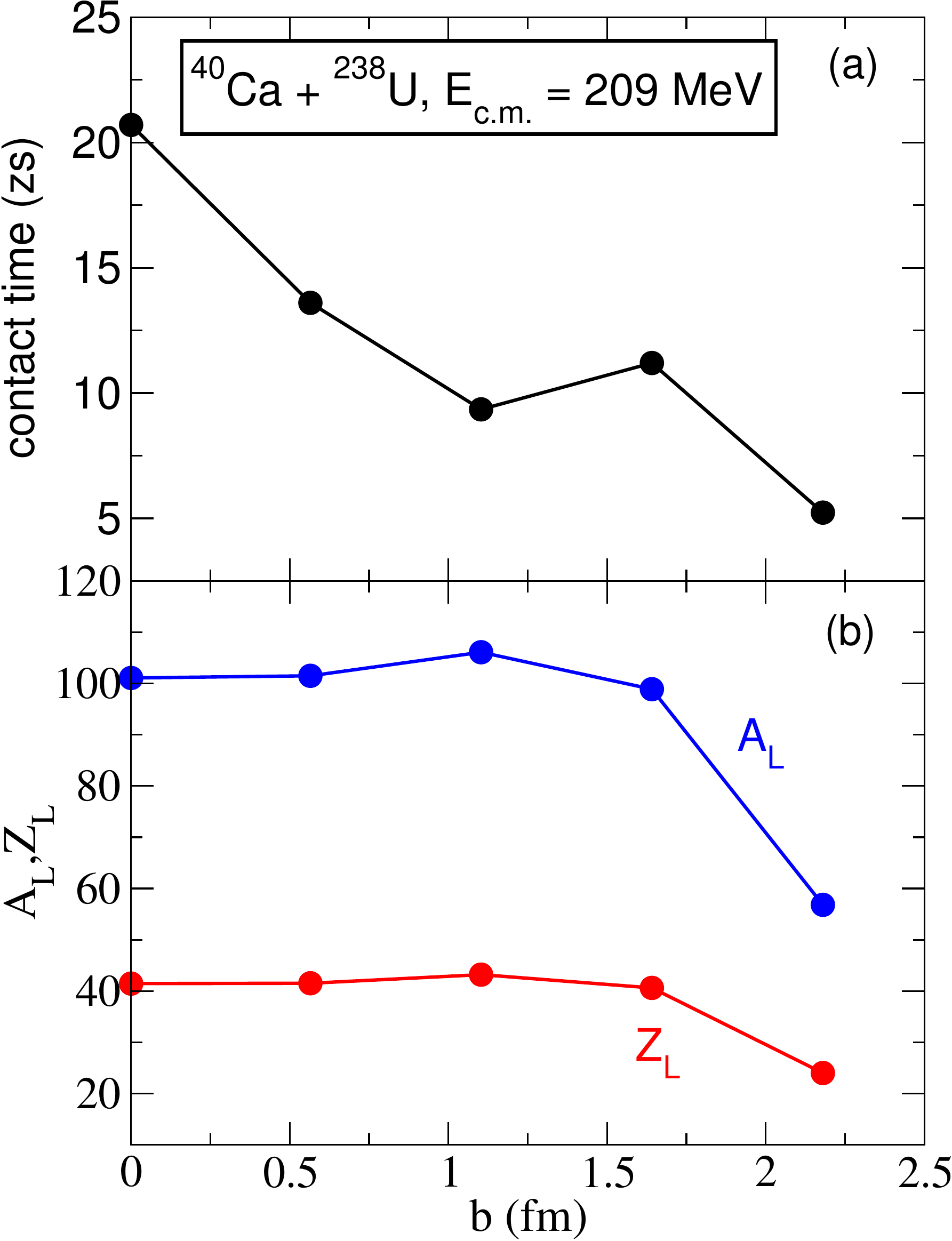}
	\caption{\protect(Color online) (a) contact time and (b) mass and
		charge of the light fragment as a function of impact parameter.}
	\label{fig:40Ca_stack_2}
\end{figure}

Figure~\ref{fig:Ca_stack_1}a displays the contact time as a function of the 
ratio of the center-of-mass energy $E_{c.m.}$ with the frozen Hartree-Fock barrier for central collisions. 
This barrier is calculated for collisions with the side of $^{238}$U~\cite{simenel2013b}.
Two fragments are always observed in the exit channel 
up to c.m. energies $\sim10\%$ above the barrier. 
Globally, the contact time increases with energy
and reaches a maximum of 32~zs at c.m. energy $\sim10\%$ above the barrier. 
TDHF calculations carried out at higher energy ($E_{\mathrm{c.m.}}\ge 223$~MeV) 
show {\it one} fragment at the end of the calculation. 
In this case, contact times exceed $35$~zs, which is interpreted as possible fusion reactions 
leading to the formation of a compound nucleus. 
The effect of a finite impact parameter $b$ is to reduce this contact time as shown in Fig.\ref{fig:40Ca_stack_2}a.
The above observations  are consistent with the fact that contact time increases 
as matter overlap between the fragments 
at the distance of closest approach increases. 
However, we also observe a plateau at $\sim20$~zs above $1.05V_B$ in Fig.~\ref{fig:Ca_stack_1}a 
which cannot be explained with such simple considerations. 


These contact times are long enough to enable the transfer of a large number of nucleons as shown in 
Fig.~\ref{fig:Ca_stack_1}b where the masses $A_\mathrm{L}$ and charges $Z_\mathrm{L}$ of
the light fragment are plotted. 
A  plateau is again observed 
for energies of $\sim5-10\%$ above $V_B$.
This corresponds to a light fragment with $Z_L\simeq40-42$ and $A_L\simeq100-107$. 
Varying the impact parameter up to $\sim2$~fm does not alter these observations as shown in Fig.~\ref{fig:40Ca_stack_2}b.
The root of this behavior may be due to the fact that  Zr isotopes ($Z=40$) in the mass range $100-112$ are strongly bound
with a large prolate deformation around $\beta_2=0.42$~\cite{lalazissis1999,blazkiewicz2005,hwang2006}.
Due to shell effects, these configurations may be energetically favorable during the QF dynamics~\cite{itkis2004,nishio2008,kozulin2014}.
A similar effect is observed in TDHF calculations of collisions with the tip of $^{238}$U 
which favors the formation of fragments in the vicinity of the doubly magic $^{208}$Pb nucleus~\cite{wakhle2014}. 

\begin{figure}[!htb]
	\includegraphics*[width=8.6cm]{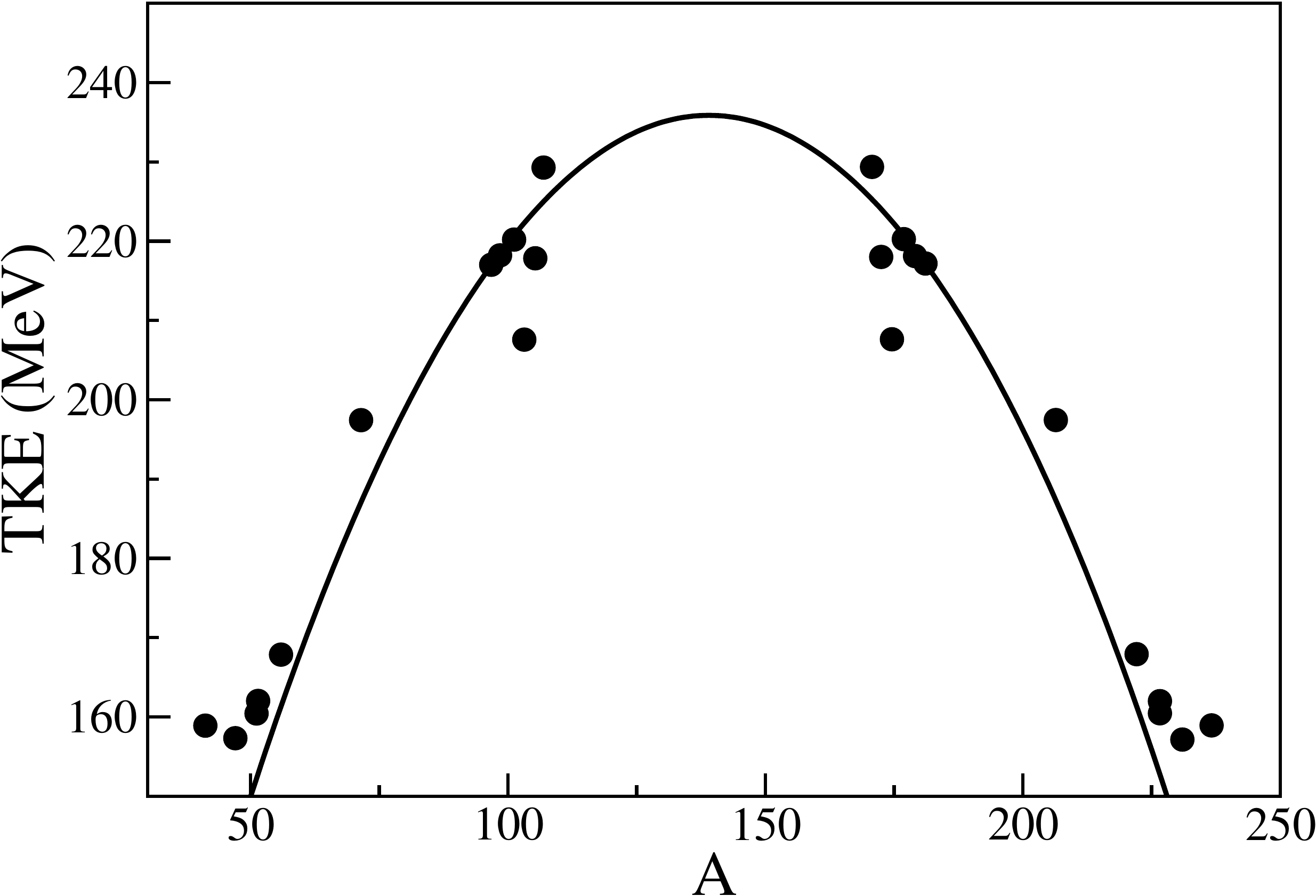}
	\caption{\protect TKE of the fragments formed in $^{40}$Ca+$^{238}$U central collisions at $E_{c.m.}/V_B=1.0-1.1$.
		The solid line represents TKE values based on the Viola formula~\cite{viola1985}.}
	\label{fig:TKE}
\end{figure}

The quasi-fission  contact times are also long enough to enable a strong damping of the initial relative kinetic energy due to dissipation mechanisms. 
Experimentally, the measured total kinetic energy (TKE) of the quasi-fission fragments 
in  $^{40,48}$Ca+$^{238}$U reactions is in relatively good agreement with the Viola systematics~\cite{toke1985,nishio2012}.
The TDHF approach contains one-body dissipation mechanisms which are dominant at near-barrier energy.
It can then be used to predict the final TKE of the fragments. 
The TKE of the fragments formed in $^{40}$Ca+$^{238}$U have been computed for a range of central collisions up to $10\%$ above the barrier.  
Figure~\ref{fig:TKE} shows that the TDHF predictions of TKE are in excellent agreement with the Viola systematics~\cite{viola1985}. 
This indicates that collisions leading to quasi-fission are indeed fully damped, leading to excited fragments. 

The excitation energy and, in particular, its repartition between the fragments 
also provide important information on the dissipative nature of the reaction mechanisms~\cite{toke1992,klein1997,nishinaka2004,schmidt2011}. 
Recently, we have developed an extension to TDHF theory via the use
of a density constraint to calculate fragment excitation energy of {\it each fragment} directly from the
TDHF time evolution~\cite{umar2009a}.
This gives us a new information on the repartition of the excitation energy between the heavy and light fragments 
which is not available in standard TDHF calculations. 
In Fig.~\ref{fig:Ca_stack_1}c we show the excitation energies of the light and heavy fragments. 
For $^{40}$Ca+$^{238}$U at 5 to $10\%$ above the barrier, we find excitation energies
which are approximately constant with $E^{*(TDHF)}_H\simeq60$~MeV for the heavy fragment and
$E^{*(TDHF)}_L\simeq40$~MeV for the light fragment.
In the assumption that the system is in statistical equilibrium, we expect the same excitation energy per nucleon in both fragments, i.e., 
$E^{*(eq)}_i\simeq E^{*(eq)}_{tot}\frac{A_i}{A_{tot}}$. 
With a total excitation energy $E^{*(eq)}_{tot}\simeq100$~MeV and fragment masses $A_L\simeq100$ and $A_H\simeq178$ (see Fig.~\ref{fig:Ca_stack_1}b), 
this assumption would give $E^{*(eq)}_H\simeq64$~MeV and $E^{*(eq)}_L\simeq36$~MeV.
As a first approximation, the statistical equilibrium assumption is then in reasonable agreement with the TDHF predictions. 
This indicates that the nucleon transfer mechanism for $^{40}$Ca+$^{238}$U central collisions 
in this energy range is of dissipative nature, and not due to shape fluctuations of the fragments. 

We have performed similar TDHF calculations for the more neutron-rich system
$^{48}$Ca+$^{238}$U, with the purpose of investigating the role of neutron to proton ratio $N/Z$ asymmetry of the colliding partners. 
Indeed, unlike $^{40}$Ca ($N/Z=1$), the more neutron rich $^{48}$Ca nucleus has an $N/Z=1.4$ which is close to the $^{238}$U one ($N/Z\simeq1.6$).
As shown in Fig.~\ref{fig:Ca_stack_1}a-c, the TDHF predictions with $^{48}$Ca are dramatically different as compared 
to the $^{40}$Ca+$^{238}$U system: the quasi-fission region,
as evidenced by long contact time and large mass transfer, is confined
to a very narrow energy window with $E_{\mathrm{c.m.}}/V_B\simeq1.03-1.04$.
Above these energies, large contact times exceeding 35~zs are found with $^{48}$Ca. 
The onset for fusion occurs then at much lower energy with $^{48}$Ca than with $^{40}$Ca.

This difference between both reactions could be due to the total neutron number 
and/or to the different initial $N/Z$ asymmetries. 
Experimental investigations with similar projectiles at near barrier energies have concluded 
that the variation of quasi-fission must be related to the properties in the entrance channel, 
rather than properties of the composite system~\cite{simenel2012b}.
It has also been argued in the same work that the hindrance of quasi-fission with $^{48}$Ca
is due to the fact that it is a doubly-magic nucleus and that it essentially keeps its magicity 
when it collides with a target of similar $N/Z$. 
Indeed, spherical shells are expected to result in so-called ''cold valleys'' 
in the potential energy surface leading to the compact compound nuclei~\cite{sandulescu1976,fazio2005,aritomo2006}. 
Fusion through these valleys may also be favored because energy dissipation should be weaker, 
allowing greater interpenetration before the initial kinetic energy is dissipated~\cite{hinde2005,armbruster2000}.
This last point is supported by the fact that both exit fragments have similar excitation energies 
in the $^{48}$Ca+$^{238}$U reaction (see Fig.~\ref{fig:Ca_stack_1}c).
Indeed, the fact that $E^*_i$ is not proportional to $A_i$ indicates that the system is out of thermal equilibrium, 
and that the transfer is not only dissipative, but could be affected by shape fluctuations.
On the contrary, $^{40}$Ca, which is also a doubly magic nucleus but with a smaller $N/Z$,
encounters a rapid $N/Z$ equilibration in the early stage of the collision, modifying its identity.
As a result, $^{40}$Ca essentially behaves as a non-magic nucleus, i.e., with more quasi-fission. 

In summary, we have done a comparative study of QF for the $^{40,48}$Ca+$^{238}$U systems
using the microscopic TDHF theory. 
Both systems exhibit fully damped events with contact time 
(up to $\sim30$~zs) and large mass transfer typical to QF.
However, the $^{48}$Ca+$^{238}$U
system shows considerably less QF in comparison to the $^{40}$Ca+$^{238}$U system.
This elucidates the success of SHE synthesis with $^{48}$Ca beams. 
The origin of the difference between both reactions is attributed 
to a longer survival of the magicity of $^{48}$Ca in the collision process which reduces dissipation mechanisms. 
This scenario is supported by the  new microscopic calculations of the repartition of the excitation energy 
between the fragments which indicate that the compound system formed in $^{48}$Ca+$^{238}$U is out of thermal equilibrium. 
In contrast, collisions with $^{40}$Ca encounter a rapid $N/Z$ equilibration resulting in dissipative transfer. 

We thank D. J. Hinde for useful discussion. 
This work has been supported by the U.S. Department of Energy under grant No.
DE-FG02-96ER40975 with Vanderbilt University and by the
Australian Research Council Grant No. FT120100760.


\bibliography{VU_bibtex_master.bib}


\end{document}